\documentclass[sigconf,screen]{acmart}

\AtBeginDocument{%
  \providecommand\BibTeX{{%
    \normalfont B\kern-0.5em{\scshape i\kern-0.25em b}\kern-0.8em\TeX}}}

\setcopyright{acmcopyright}
\copyrightyear{2023}
\acmYear{2023}
\acmDOI{XXXXXXX.XXXXXXX}

\acmConference[ESEC/FSE 2023]{The 31st ACM Joint European Software Engineering Conference and Symposium on the Foundations of Software Engineering}{11 - 17 November, 2023}{San Francisco, USA}

\acmPrice{15.00}
\acmISBN{978-1-4503-XXXX-X/18/06}

\usepackage{amsmath,amsfonts}
\usepackage{algorithmic}
\usepackage{graphicx}
\usepackage{textcomp}
\usepackage{ulem}
\usepackage{xcolor}
\usepackage{color, colortbl}
\usepackage{balance}
\usepackage{url}

\usepackage{xspace}

\newcommand{\etal}{\hbox{et al.}\xspace}
\newcommand{\eg}{\hbox{e.g.}\xspace}
\newcommand{\ie}{\hbox{i.e.}\xspace}

\newcommand{\ourmethod}{\textit{SSQR}\xspace}

\usepackage{mathrsfs}

\newcommand{\codecomment}[1]{\hfill$\triangleright$ {#1}}
\usepackage{multirow}
\usepackage{booktabs}
\usepackage{listings}
\usepackage{graphicx}
\usepackage{subcaption}
\usepackage{cleveref}
\usepackage{setspace}
\usepackage{threeparttable}
\usepackage[ruled]{algorithm2e}
\begin{document}

\title{Self-Supervised Query Reformulation for Code Search}

\settopmatter{authorsperrow=4}
\author{Yuetian Mao}
\authornote{Both authors contributed equally to this research.}
\affiliation{%
  \institution{Shanghai Jiao Tong University}
  \city{Shanghai}
  \country{China}
}
\email{mytkeroro@sjtu.edu.cn}

\author{Chengcheng Wan}
\authornotemark[1]
\affiliation{%
  \institution{East China Normal University}
  \city{Shanghai}
  \country{China}
}
\email{wancc1995@gmail.com}

\author{Yuze Jiang}
\affiliation{%
  \institution{Shanghai Jiao Tong University}
  \city{Shanghai}
  \country{China}
}
\email{jyz-1201@sjtu.edu.cn}

\author{Xiaodong Gu}
\authornote{Xiaodong Gu is the corresponding author.}
\affiliation{%
  \institution{Shanghai Jiao Tong University}
  \city{Shanghai}
  \country{China}
}
\email{xiaodong.gu@sjtu.edu.cn}


\begin{abstract}
Automatic query reformulation is a widely utilized technology for enriching user requirements and enhancing the outcomes of code search. It can be conceptualized as a machine translation task, wherein the objective is to rephrase a given query into a more comprehensive alternative. While showing promising results, training such a model typically requires a large parallel corpus of query pairs (\ie, the original query and a reformulated query) that are confidential and unpublished by online code search engines. This restricts its practicality in software development processes. In this paper, we propose \ourmethod, a self-supervised query reformulation method that does not rely on any parallel query corpus. Inspired by pre-trained models, \ourmethod treats query reformulation as a masked language modeling task conducted on an extensive unannotated corpus of queries. \ourmethod extends T5 (a sequence-to-sequence model based on Transformer) with a new pre-training objective named \textit{corrupted query completion} (CQC), which randomly masks words within a complete query and trains T5 to predict the masked content. Subsequently, for a given query to be reformulated, \ourmethod identifies potential locations for expansion and leverages the pre-trained T5 model to generate appropriate content to fill these gaps. The selection of expansions is then based on the information gain associated with each candidate. Evaluation results demonstrate that \textit{our method} outperforms unsupervised baselines significantly and achieves competitive performance compared to supervised methods.
\end{abstract}



\maketitle

\section{Introduction}

Searching through a vast repository of source code has been an indispensable activity for developers throughout the software development process~\cite{XiaBLKHX17}. The objective of code search is to retrieve and reuse code snippets from existing projects that align with a developer's intent expressed as a natural language query \cite{YanYCSJ20}.
However, it has been observed that developers often struggle to articulate their information needs optimally when submitting queries~\cite{KoMCA06, eberhart2022generating}. 
This difficulty may arise from factors such as inconsistent terminology used in the query or a limited understanding of the specific domain in which information is sought. 
Developers may constantly reformulate their queries until the queries reflect their real query intention and retrieve the most relevant code snippets. 
Studies~\cite{sequer} have shown that in Stack Overflow, approximately 24.62\% of queries on Stack Overflow have undergone reformulation. Moreover, developers, on average, reformulate their queries 1.46 times before selecting a particular result to view.

One common solution to this problem is automatic query reformulation, namely, rephrasing a given query into a more comprehensive alternative~\cite{HaiducBMOLM13, liu2022formulate}. 
A natural first way to accomplish this objective is to replace words in a query with synonyms based on external knowledge such as WordNet and thesauri \cite{sesimilarwords, SWordNet, lusearch, li2022cooperative}. However, this methodology restricts the expansion to the word level. Besides, gathering and maintaining domain knowledge is usually costly. The knowledge base might always lag behind the fast-growing code corpora. 
There have been other attempts that consider pseudo-relevance feedback, i.e., emerging keywords in the initial search results \cite{nlp2api, pseudo, huang2017query, zhu2022lol}. They search for an initial set of results using the original query, select new keywords from the top \textit{k} results using TF-IDF weighting, and finally expand the original query with the emerging keywords. 
Nevertheless, despite expanding queries at a word level, this approach also has a risk of expanding queries with noisy words. Hence, the expanded query can be semantically irrelevant to the original one.

In recent years, driven by the prevalence of deep learning, researchers seek the idea of casting query reformulation as a machine translation task: the original query is taken as input to a neural sequence-to-sequence model and is translated into a more comprehensive alternative~\cite{sequer}. 
Despite showing substantial gains, such models require to be trained on a large-scale parallel corpus of query pairs (\ie, the original query and a reformulated query). Unfortunately, acquiring large query pairs is infeasible given that real-world search engines (\eg, Google and Stack Overflow) do not publicly release the evolution of queries. For example, the state-of-the-art method SEQUER \cite{sequer} relies on a confidential parallel dataset that cannot (likely to be impossible) be accessed by external researchers. Replicating the performance of SEQUER becomes challenging or even impossible for those who lack access to such privileged datasets. This lack of replicability hampers the wider adoption and evaluation of the method by the research community.

In this paper, we present \ourmethod, a self-supervised query reformulation method that achieves competitive performance to the state-of-the-art supervised approaches, while not relying on the availability of parallel query data for supervision. Inspired by the pre-trained models, \ourmethod automatically acquires the supervision of query expansion through self-supervised training on a large-scale corpus of code comments.  
Specifically, we design a new pre-training objective called \textit{corrupted query completion} (CQC) to simulate the query expansion process. CQC masks keywords in long, comprehensive queries and asks the model to predict the missing contents. In such a way, the trained model is encouraged to expand incomplete queries with keywords.
\ourmethod leverages T5~\cite{t5}, the state-of-the-art language model for code. The methodology of \ourmethod involves a two-step process. Firstly, T5 is pre-trained using the CQC objective on a vast unannotated corpus of queries. This pre-training phase aims to equip T5 with the ability to predict masked content within queries.
When presented with a query to be reformulated, \ourmethod enumerates potential positions within the query that can be expanded. It then utilizes the pre-trained T5 model to generate appropriate content to fill these identified positions. Subsequently, \ourmethod employs an information gain criterion to select the expansion positions that contribute the most valuable information to the original query, resulting in the reformulated query.

We evaluate \ourmethod on two search engines through both automatic and human evaluations, and compare with state-of-the-art approaches, including SEQUER~\cite{sequer}, NLP2API~\cite{nlp2api}, LuSearch~\cite{lusearch}, and GooglePS~\cite{GooglePS}. Experimental results show that \ourmethod improves the MRR score by over 50\% compared with the unsupervised baselines and gains competitive performance over the fully-supervised approach. 
Human evaluation reveals that our approach can generate more natural and informative queries, with improvements of 19.31\% and 26.35\% to the original queries, respectively. 

Our contributions are summarized as follows:
\begin{itemize}
    \item To the best of our knowledge, \ourmethod is the first self-supervised query reformulation approach, which does not rely on a parallel corpus of reformulations.
    \item We propose a novel information gain criterion to select the pertinent expansion positions that contribute the most valuable information to the original query.
    \item We perform automatic and human evaluations on the proposed method. 
    Quantitative and qualitative results show significant improvements over the state-of-the-art approaches.     
\end{itemize}

\section{Background}
\subsection{Code Search}
Code search is a technology to retrieve and reuse code from pre-existing projects~\cite{deepcs,ChaiZSG22,YanYCSJ20}. 
Similar to general-purpose search engines, developers often encounter challenges when attempting to implement specific tasks. In such scenarios, they can leverage a code search engine by submitting a natural language query. The search engine then traverses an extensive repository of code snippets collected from various projects, identifying code that is semantically relevant to the given query. Code search can be broadly classified into two categories: search within the context of a specific project or as an open search across multiple projects. The search results may include individual code snippets, functions, or entire projects.

\subsection{Query Reformulation}
\begin{figure}
 \centering
  \includegraphics[width=0.8\linewidth, trim=30 10 30 10 clip]{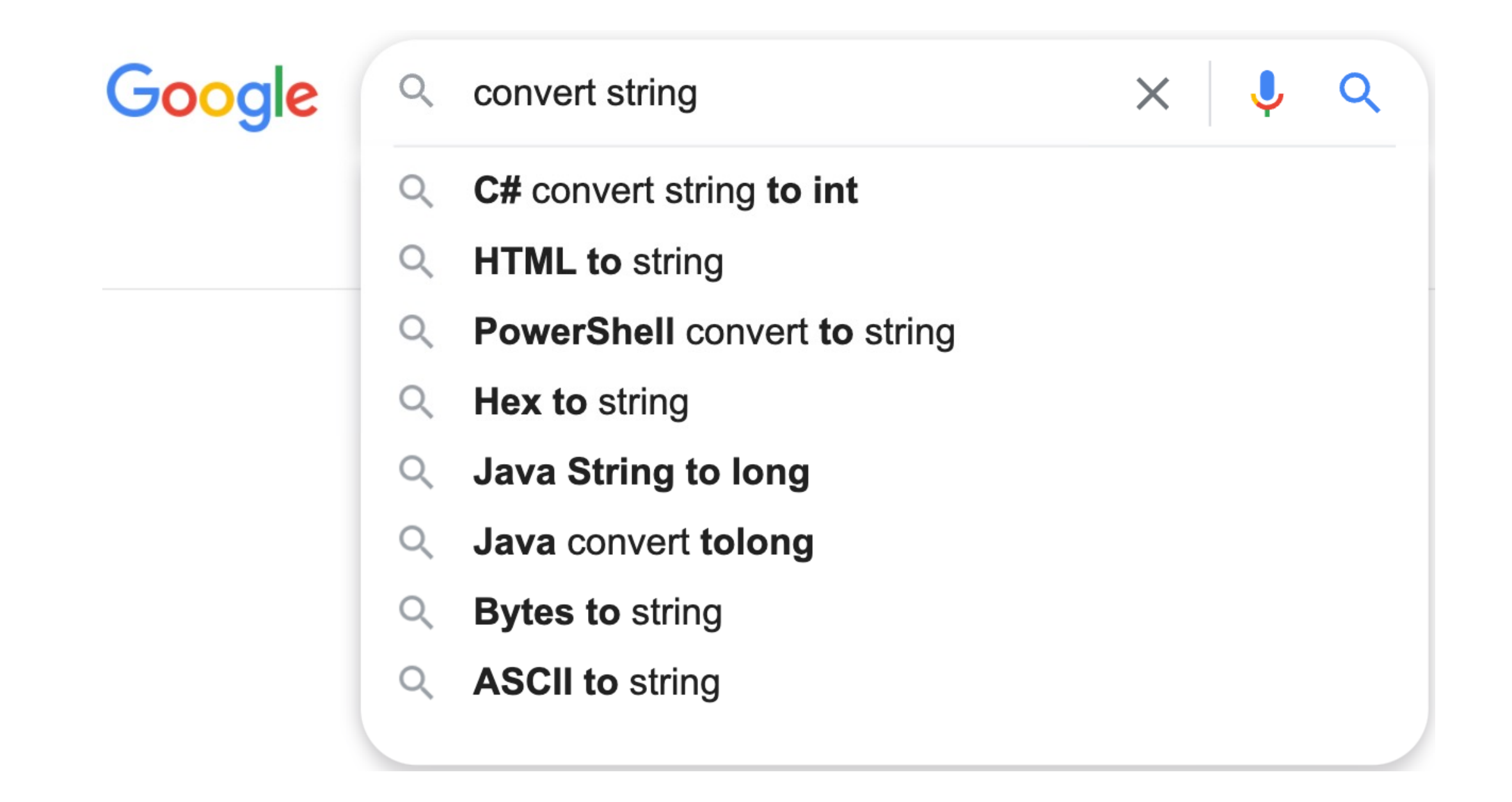}
  \caption{An example of Google query reformulation.}
  \label{fig:QueryReformulationExample}
\end{figure}

\begin{figure*}[ht]
 \centering
  \includegraphics[width=0.8\linewidth, trim=0 10 0 5, clip]{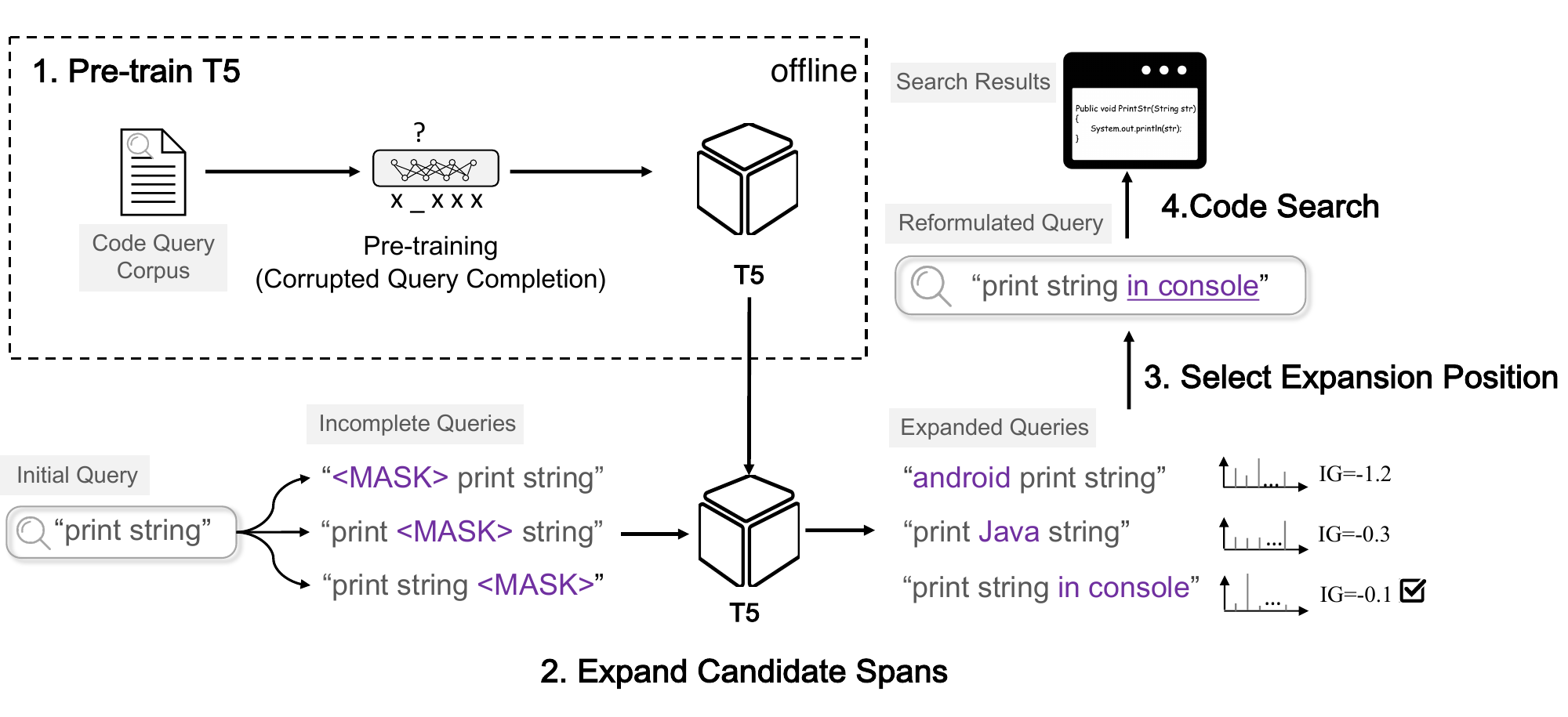}
  \caption{An illustration of the main pipeline}
  \label{fig:framework}
\end{figure*}

 \begin{figure}[h]
 \centering
  \includegraphics[width=0.85\linewidth, trim=10 10 10 40 clip]{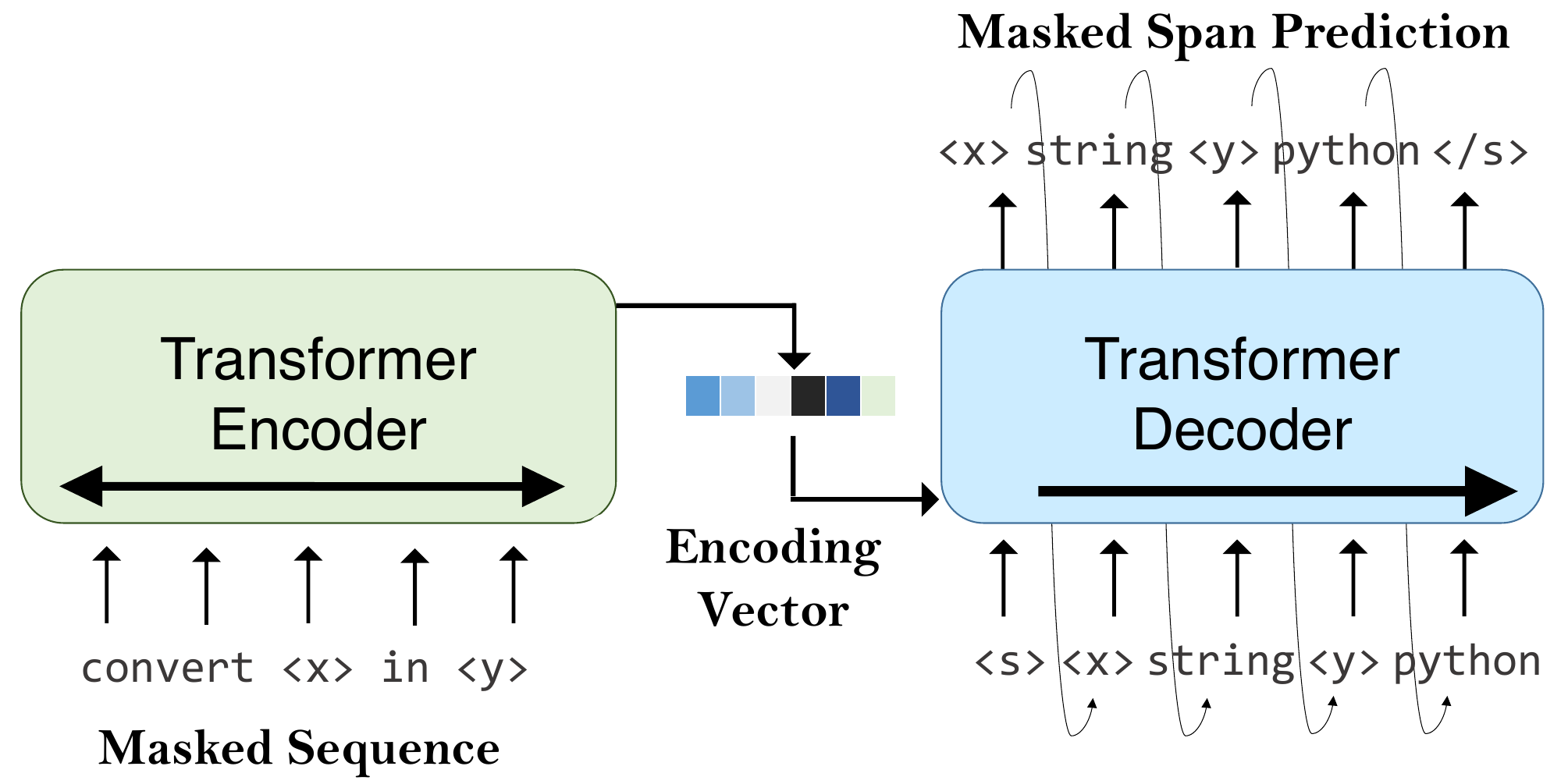}
  \caption{Illustration of T5}
  \label{fig:T5Structure}
\end{figure}

Query reformulation provides an effective way to enhance the performance of search engines~\cite{zhu2022lol,li2022cooperative,chen2021towards}. 
The quality of queries is often a bottleneck of search experience in web search~\cite{chen2021towards}. This is because the initial query entered by the user is often short, generic, and ambiguous. Therefore, the search results could hardly meet the specific intents of the user. This requires the user to revise his query through multiple rounds. Query reformulation is a technology that reformulates user's queries into more concrete and comprehensive alternatives~\cite{huang2009analyzing}.
Figure~\ref{fig:QueryReformulationExample} shows an example of query reformulation in Google search engine. When a user enters the query ``convert string'' in the search box, there may exist multiple possible intents, such as ``convert something to a string'' or ``convert a string to something''. Additionally, the specific programming language for implementing the conversion function is not specified. In such cases, conventional search engines like Google face challenges in accurately determining the user's true intent. 

To address this issue, search engines often employ tools like the Google Prediction Service (GooglePS). GooglePS automatically suggests multiple reformulations of the original query. These reformulations provide alternative options that the user can consider to refine their search. By presenting a range of reformulations, users can narrow down their search target by selecting the most relevant reformulation that aligns with their intended query. This process helps users in finding more precise and tailored search results.

Query reformulation broadly encompasses various techniques, including query expansion, reduction, and replacement~\cite{jansen2009patterns}. While query expansion involves augmenting the original query with additional information, such as synonyms and related entities, to enhance its content, query reduction focuses on eliminating ambiguous or inaccurate expressions. Query replacement, on the other hand, involves substituting incorrect or uncommon keywords in the original query with more commonly used and precise terms. Among these types, query expansion constitutes the predominant approach, accounting for approximately 80\% of real-world search scenarios~\cite{sadowski2015developers}.

\subsection{Self-Supervised Learning and Pre-trained Models}

Supervised learning is a class of machine learning methods that train algorithms to classify data or predict outcomes by leveraging labeled datasets.
It is known to be expensive in manual labeling, and the bottleneck of data annotation further causes generalization errors~\cite{jakubovitz2019generalization}, spurious correlations~\cite{kronmal1993spurious}, and adversarial attacks~\cite{madry2017towards}. 
Self-supervised learning alleviates these limitations by automatically mining supervision signals from large-scale unsupervised data using auxiliary tasks \cite{Self-supervised-Learning}. This enables a neural network model to learn rich representations without the need for manual labeling \cite{Yang0S22}. For example, the cloze test masks words in an input sentence and asks the model to predict the original words. In this way, the model can learn the semantic representations of sentences from large unlabeled text corpora. 

Pre-trained language models (PLMs) such as BERT~\cite{BERT}, GPT~\cite{GPT}, and T5~\cite{t5} are the most typical self-supervised learning technology. A PLM aims to learn language's generic representations on a large unlabeled corpus and then transfer them to specific tasks through fine-tuning on labeled task-specific datasets. This requires the model to create self-supervised learning objectives from the unlabeled corpora. 
Take the Text-to-Text Transfer Transformer (T5)~\cite{t5} in Figure~\ref{fig:T5Structure} as an example. T5 employs the Transformer~\cite{vaswani2017attention} architecture where an encoder accepts a text as input and outputs the encoded vector. 
A decoder generates the target sequence based on the encodings. To efficiently learn the text representations, T5 designs three self-supervised pre-training tasks, namely, masked span prediction, masked language modeling, and corrupted sentence reconstruction.
By pre-training on large-scale text corpora, T5 achieves state-of-the-art performance in a variety of NLP tasks, such as sentence acceptability judgment~\cite{zomer2021beyond}, sentiment analysis~\cite{pipalia2020comparative}, paraphrasing similarity calculation~\cite{nighojkar2021improving}, and question answering~\cite{jiang2021can}.


\section{Method}

The primary focus of this paper is on query expansion, the most typical (accounting for 80\%) technique for query reformulation. Query expansion aims to insert key phrases into a query thereby making it more specific and comprehensive. 
Essentially, query expansion addresses a \textit{pinpoint-then-expand} problem, wherein the goal is to identify potential information gaps within a given query and generate a set of keywords to fill those gaps. 

Inspired by the masked language modeling (MLM) task introduced by pre-trained models like BERT~\cite{BERT}, our proposed method adopts a self-supervised idea. Specifically, we mask keywords within complete code search queries and train a model to accurately predict and recover the masked information. This allows the model to learn the underlying patterns and relationships within the queries, enabling it to generate meaningful expansions for query reformulation.

\begin{figure*}[t]
 \centering
  \includegraphics[width=1.0\linewidth]{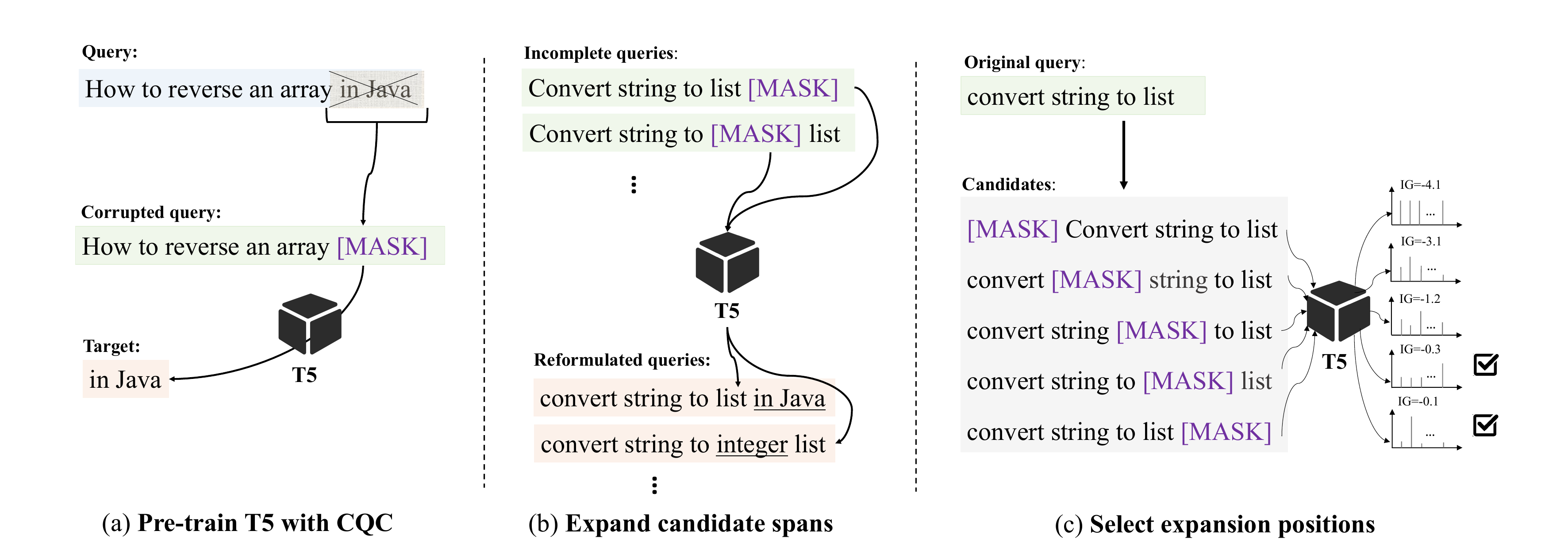}
  \caption{A working example of each expansion step}
  \label{fig:example}
\end{figure*}

\subsection{Overview}
Figure~\ref{fig:framework} shows the main framework as well as the usage scenario of our method. The pipeline involves two main phases: an offline pre-training phase and an online expansion phase. 
During the pre-training phase, \ourmethod continually pre-trains a PLM named T5~\cite{t5} with a newly designed \textit{corrupted query completion} task on an unlabelled corpus of long queries (\S \ref{ss:approach:pretrain}). This enables the T5 to learn how to expand incomplete queries into longer ones. 
During the runtime of \ourmethod, when a user presents a query for code search, \ourmethod employs a two-step process for query expansion. Firstly, it enumerates candidate positions within the query that can be expanded and utilizes the pre-trained T5 model to generate content that fills these positions (as discussed in \S \ref{ss:approach:step2}). 
Following the expansion step, \ourmethod proceeds to select the position that offer the highest information gain after the expansion (introduced in \S \ref{ss:approach:selection}). This selection process ensures that the most valuable and informative expansions are chosen, thereby enhancing the reformulated query in terms of its relevance and comprehensiveness.

Finally, once the query has been expanded, users conduct code search by selecting the most relevant reformulation that aligns with their intended query. Our approach specifically focuses on the function-level code search scenario, which involves the retrieval of relevant functions from a vast collection of code snippets spanning multiple projects.

The following sections elaborate on each step of our approach respectively. 

\subsection{Pre-training T5 with Corrupted Query Completion}
\label{ss:approach:pretrain}

We start by pre-training a PLM which can predict the missing span in a query. 
We take the state-of-the-art T5~\cite{t5} as the backbone model since it has a sequence-to-sequence architecture and is more compatible with generative tasks. Besides, T5 is specialized in predicting masked spans (\ie, a number of words). 

To enable T5 to learn how to express a query more comprehensively, we design a new pre-training objective called \textit{corrupted query completion} (CQC) using a large-scale corpus of unlabelled queries. 
Similar to the MLM objective, CQC randomly masks a span of words in the query and asks the model to predict the masked span. 
More specifically, given an original query $q=\left(w_{1}, \ldots,  w_{n}\right)$ that consists of a sequence of $n$ words, \ourmethod masks out a span of 15\%$\times$$n$ consecutive words from a randomly selected position $i$, namely, $s_{i: j}$=$\left(w_{i}, \ldots, w_{j}\right)$, and replaces it with a \texttt{[MASK]} token. 
Then, the corrupted query is taken as input to T5 which predicts the words in the masked span. 
We use the teacher-forcing strategy for pre-training. When predicting a word in the corrupted span, the context visible to the model consists of two parts: 1) the uncorrupted words in the original query, denoted as $q_{\backslash s_{i: j}}=\left(w_{1}, \ldots, w_{i-1}, w_{j+1}, \cdots, w_{n}\right)$; and 2) the ground truth words appeared before the current predicting position $w_{t}$, denoted as $w_{i: t-1}$. We pre-train the model using the cross-entropy loss, namely, minimizing
\begin{equation}
\mathcal{L}_{\mathrm{cqc}}=-\sum_{t=i}^{j} \log p\left(w_{t} \mid q_{\backslash s_{i: j}}, w_{i: t-1}\right).
\end{equation}
Figure~\ref{fig:example}(a) shows an example of the CQC task. For a query ``how to reverse an array in Java'' taken from the training corpus, the algorithm corrupts the query by replacing the modifier ``in Java'' with \texttt{[MASK]}. The corrupted query is taken as input to T5 which predicts the original masked tokens ``in Java''.


\subsection{Expanding Candidate Spans} 
\label{ss:approach:step2}
The pre-trained T5 model is then leveraged to expand queries. 
We consider a query that needs to be expanded as an incomplete query where a span of words is missing at a position (denoted as a masked token), we want the model to generate a sequence of words to fill in the span. This is exactly the problem of \emph{masked span prediction} as T5 aims to solve. Therefore, we leverage the pre-trained T5 to expand the incomplete queries. 

However, a query with $n$ words have $n$+$1$ positions for expansion. Therefore, we design a \textit{best-first} strategy: we enumerate all the $n$+$1$ positions as the masked spans, perform the CQC task, and select the top-\textit{k} positions that have the most information gain of predictions.
Specially, given a original query $q$ =$\{w_1,w_2,...,w_n\}$, \ourmethod enumerates the $n$+$1$ positions between words. For each position, it inserts a \texttt{[MASK]} token. This results in $n$+$1$ candidate masked queries. 
Each incomplete query $\tilde{q}$=[$w_1,\ldots,$\verb|[MASK]|$,\ldots,w_n$] is taken as input to the pre-trained T5.
The decoder of T5 generates a span of words $s$=[$v_1,\ldots,v_m$] for the \verb|[MASK]| token by sampling words according to the predicted probabilities. 
Finally, \ourmethod replaces the \verb|[MASK]| token with the generated span $s$, yielding the reformulated query.

Figure~\ref{fig:example}(b) shows an example. Given the first two masked queries, the T5 model generates ``in Java'' and ``integer'' for the masked tokens, respectively.
The former refers to the language used to implement the function, while the latter refers to the data type of the target data structure. The reformulated queries supplement the original queries with additional information, revealing user potential intents from different aspects. 

\subsection{Selecting Expansion Positions} 
\label{ss:approach:selection}

\begin{algorithm}[t]
    \caption{Span Expansion and Selection}
    \label{algorithm:PredictingMaskPosition}
    \LinesNumbered
    \KwIn {$q$: the input query written in natural language\;
    ~~~~~~~T5: the pre-trained T5 model\;
    ~~~~~~~~\,$k$: the number of candidate positions\;
    ~~~~~~~$m$: the maximum target length.
    }
    \KwOut {$\mathbf{Q}$: a set of candidate masked queries.}
    $w_1,...,w_n$ = tokenize($q$)\;
    $\mathbf{Q}$ = \{\}; \codecomment{initialize the query set.} \\
    \For{i = 1 to n}{
        $\tilde{q}$ = $w_1,...,w_{i}$, \texttt{[MASK]}, $w_{i+1},...,w_n$;
        ~~\codecomment{insert a}\\\hfill \texttt{[MASK]} token to the $i$-th position in $q$.\\
        $v_{1},...,v_{m}$ = T5($\tilde{q}$);~~ \codecomment{predict the content for the}\\\hfill \texttt{[MASK]} token.\\
        $IG$ = $\frac{1}{m}\sum_{j=1}^m p(v_j)\mathrm{log}p(v_j)$;~~\codecomment{calculate the}\\\hfill information gain for the expansion. \\
        $\mathbf{Q} = \mathbf{Q}\cup\langle\tilde{q},IG\rangle $\;
    }  
    Sort $\mathbf{Q}$ based on entropy in an descending order\;
    $\mathbf{Q}$= top$(\mathbf{Q},k)$;~~\codecomment{ select the top-$k$ queries.}\\
    \Return $\mathbf{Q}$
\end{algorithm}

A query with $n$ words has $n$+$1$ candidate positions for expansion, but not all of them are necessary for expansion. Hence, we must determine which positions are the most proper to be expanded. 
\ourmethod selects the top-\textit{k} candidate queries that have the most missing information in the masked span. 
The resulting expanded queries are more likely to gain information after span filling. 


The key issue here is how to measure the information gain after filling each span. In our approach, we define \emph{information gain} for a span expansion as the negative entropy over the predicted probability distribution of the generated words~\cite{jiang2015relative, luukka2011feature}. In information theory, entropy~\cite{renyi1961measures} characterizes the uncertainty of an event in a system. Suppose the probability that an event will happen follows a distribution of ($p_1,\ldots,p_n$), the entropy of the event can be computed as $-\sum_{i=1}^np_i\mathrm{log}p_i$. The lower the entropy, the more certain that the event can happen. That means the event brings more information to humans.
This can be analogized to the span prediction problem: when the probability distribution of the generated words over the vocabulary is uniform, the entropy (uncertainty) becomes high because every word is likely to be generated. By contrast, smaller entropy means that there is a greatly different likelihood of generating each word and thus the certainty of the generation is high. The lower the entropy, the higher the certainty that this span contains the word, and the more information that the expansion brings to the query.
If the span contains multiple words, we can measure the information gain of the span prediction using their average negative entropy.  

For each candidate query $\tilde{q}$, we predict a span $s$=$[v_1,...,v_m]$ using the pretrained T5 model: 
\begin{equation}
p(v_i)=T5(\tilde{q}, v_{<i}), i=1,\ldots,L
\end{equation}
where each $v_i$ denotes a sub-token in the predicted span.
Each prediction $p(v_i)$ follows a probability distribution of $p_1,\ldots,p_{|V|}$ over the vocabulary of the entire set of queries in the training corpus, indicating the likelihood of each token in the vocabulary appearing in the span. 

Our next step is to compute the information gain of each expansion using negative entropy: For each sub-token $v_i$, the information gain can be calculated by 
\begin{equation}
IG(v_i) = -H(v_i)=\sum_{v=1}^{|V|}p(v_i=v)\mathrm{log} p(v_i=v). 
\end{equation}
The higher the IG, the more certainty that the prediction is. For a predicted span $s$ = [$v_1,\ldots,v_m$] with $m$ sub-tokens, we compute the average IG of all its tokens, namely, 
\begin{equation}
IG(s) = \frac{1}{m}\sum_{i=1}^LIG(v_i). 
\end{equation}

Finally, we select the top $K$ expansions with the highest information gain and then replace the \texttt{[MASK]} token with the predicted span. The top-$k$ expansions are provided to users for choosing the most relevant one that aligns with their intention. 
The specific details of the method are summarized in Algorithm~\ref{algorithm:PredictingMaskPosition}.

Figure~\ref{fig:example}(c) shows an example query expansion. For a given query ``convert string to list'' to be expanded, \ourmethod firstly enumerates five expansion positions of the original query, inserting a \texttt{[MASK]} token into each one. Next, the pre-trained T5 model takes these candidate queries as input and calculates the information gain from the prediction for these candidate queries. Finally, \ourmethod recommends the top-2 masked queries (here \textit{k}=2) with the highest information gain (\ie, minimum entropy values) for users to choose.

\section{Experimental Setup}

\subsection{Research Questions}
We evaluate the performance of \ourmethod in query reformulation through both automatic and human studies. We further explore the impact of different configurations on performance. 
In specific, we address the following research questions:

\begin{itemize}
    \item \textbf{RQ1: How effective is \ourmethod in query reformulation for code search?}
    
    We apply query reformulation to Lucene and CodeBERT based code search engines and compare the search accuracy before and after query reformulation by various approaches.
    
    \item \textbf{RQ2: Whether the queries reformulated by \ourmethod are more contentful and easy to understand?}
    
    In addition to the automatic evaluation of code search performance, we also want to assess the intrinsic quality of the reformulated queries. To this end, we perform a human study to assess whether the reformulated queries contain more information than the original ones and meanwhile conform to human reading habits. 
    
    \item \textbf{RQ3: How do different configurations impact the performance of \ourmethod?}

    To obtain a better insight into \ourmethod, we investigate the performance of \ourmethod under different configurations. We firstly investigate the effect of different positioning strategies, i.e., what is the best criteria to select the expansion position in the original query. We are also interested in the number of expansions for each query. 

    
\end{itemize}

\subsection{Datasets}
We pre-train, fine-tune, and test all models using two large code search corpora: CODEnn~\cite{deepcs} and CodeXGLUE~\cite{CodeXGlue}. They are non-overlapping and thus alleviate the duplicated code issue between pre-training and downstream tasks~\cite{Allamanis19}.

\smallskip\textbf{Dataset for Pre-training}. We pre-train T5 using code comments from the large-scale CODEnn dataset. CODEnn has been specifically processed for code search. 
Compared to CodeSearchNet~\cite{CodeSearchNet}, this dataset has a much larger volume (i.e., more than 10 million queries). We take the first 1 million for pre-training.

\smallskip\textbf{Dataset for Code Search}. We use the code search dataset of CodeXGLUE which provides queries and the corresponding code segments from multiple projects. 
In this dataset, each record has five attributes, including the code segment (in Python), repository URL, code tokens, doc string (i.e., NL description of the function), and the index of the code segment. 
We split the original dataset into training and test sets, with 251,820 and 19,210 samples respectively.
The training set is used to fine-tune the CodeBERT search engine, while the test set is used as the search pool from which a search engine retrieves code. All queries in the test phase are tests on the same pool.
The statistics of our datasets are summarized in Table~\ref{tab:dataset}.

\begin{table}[!t]
\caption{\rmfamily Statistics of Datasets}
\rmfamily
\centering
\label{tab:dataset}
\begin{threeparttable}
\begin{tabular}{lcc}
\toprule
\textbf{Stage} & \textbf{Dataset} & \textbf{\# of Samples}   \\ \bottomrule
Pre-training                 & CODEnn~\cite{deepcs} & 1,000,000  \\
Fine-tuning                 & CodeXGLUE~\cite{CodeXGlue} &  251,820 \\
Search    &   CodeXGLUE~\cite{CodeXGlue} & 19,210  \\ \bottomrule
\end{tabular}
\end{threeparttable}
\vspace{-3mm}
\end{table}

\subsection{Implementation Details}
\textbf{Implementation of the pre-trained model}. Our model is implemented based on \verb|T5-base| from the open-source collection of HuggingFace~\cite{t5base}. We use the default tokenizer and input-output lengths.
Since HuggingFace does not provide an official PyTorch script for T5 pre-training, we implement the pre-training script based on the PyTorch Lightening framework~\cite{PyTorch-Lightning}. We initialize T5 with the default checkpoint provided by Huggingface and continually pre-train it with our proposed CQC task. The pre-training takes 3 epochs with a learning rate of 1e-3. The batch size is set to 32 in all experiments. We set $m$ and $k$ in Algorithm 1 to 10 and 3 respectively.
Since T5 is non-deterministic, \ourmethod can generate different queries for an input query at each run. To guarantee the same output at each run, we fix the random seeds to 101 and reload the same \emph{state-dict} of T5.

\smallskip\textbf{Implementation of the search engines}. We experiment under two search engines based on CodeBERT~\cite{feng2020codebert} and Lucene~\cite{Lucene}. 

1) \textit{A CodeBERT-based search engine}: As our approach is built on pre-trained models, we first verify the effectiveness on a pre-training based search engine. Specifically, we test our approach on the default search engine by CodeBERT. We reuse the implementation of the code search (\ie, Text-Code) task in CodeXGLUE~\cite{CodeXGlue}. Then, we fine-tune the CodeBERT-base checkpoint on a training set from CodeXGLUE for 2 epochs with a constant learning rate of 5e-5. 

2) \textit{The Lucene search engine}: Besides the pre-training based search engine, we also test our approach on a classic search engine named Lucene~\cite{Lucene}. Lucene is a keyword-based search library that is widely adapted to a variety of code search engines and platforms~\cite{solr, indextank, gormley2015elasticsearch}. We implement the Lucene search engine based on the Lucene core in Java. We extract the code segments from the test dataset from CodeXGLUE, parse them by Lucene's StandardAnalyzer, and build their indexes. 

We train all models on a Linux server with Ubuntu 18.04.1 and a GPU of Nvidia GeForce RTX 2080 Ti.

\subsection{Baselines}
We compare our method with the state-of-the-art query reformulation approaches, including a supervised method called SEQUER, and unsupervised methods such as NLP2API, LuSearch, and GooglePS. 
\begin{itemize}
\item[1)] \textbf{Supervised}~\cite{sequer}: a supervised learning approach for query reformulation named SEQUER. SEQUER leverages Transformer~\cite{vaswani2017attention} to learn the sequence-to-sequence mapping between the original and the reformulated queries. The method relies on a confidential parallel dataset of query evolution logs provided by Stack Overflow. The dataset contains internal HTTP requests processed by Stack Overflow’s web servers within one year. 
\item[2)] \textbf{NLP2API}~\cite{nlp2api}: a feedback based approach that expands query with recommended APIs.
NLP2API automatically identifies relevant API classes collected from Stack Overflow using keywords in the initial search results and then expands the query with these API classes.
\item[3)] \textbf{LuSearch}~\cite{lusearch}: a knowledge-based approach that expands a query with synonyms in WordNet~\cite{WordNet}. The reformulated queries by LuSearch are based on Lucene's structural syntax, which are too long and contain too many Lucene-specific keywords. 
Due to the constraint on the input length of T5 model (\ie, 512 tokens), we keep the synonyms and remove keywords about attribute names in Lucene such as ``methbody'' and ``methname''.
\item[4)] \textbf{GooglePS}~\cite{GooglePS}: the Google query prediction service that gives real-time suggestions on query reformulation. We directly enter test queries into the Google search box and manually collect the reformulated queries in RQ2 and the case study. We do not compare our method with GooglePS in RQ1 because its search API is unavailable to us for processing a large number of queries. Besides, our baseline model has demonstrated a great improvement over it in terms of MRR~\cite{sequer}.

\end{itemize}

\subsection{Evaluation Metrics}
The ultimate goal of query reformulation is to enhance search accuracy by using the reformulated queries. In our experiments, we first evaluate the search accuracy measured by the widely used mean reciprocal rank (MRR). MRR is defined as the average of the reciprocal ranks (\ie, the multiplicative inverse of the target post’s rank) of the search results for all the queries, namely,
\begin{equation}
MRR=\frac{1}{Q} \sum_{i=1}^{|Q|} \frac{1}{rank_{i}}
\end{equation}
where $Q$ refers to a set of queries and  $rank_{i}$ stands for the position of the first relevant document for the $i$-th query. A higher MRR indicates better search performance. 

Besides the indirect criteria in search performance, query reformulation also aims to help users write more precise and high-quality queries. Therefore, we further define two metrics to measure the intrinsic quality of the reformulated queries:
\begin{itemize}
\item \textit{Informativeness} measures how much information a query contains that contributes to code search. We use this metric to evaluate how much information gain the reformulation brings to the original query.
\item \textit{Naturalness} measures how well a query is grammatically correct and follows human reading habits. By using this metric, we want the reformulation to be semantically coherent with the original query.
\end{itemize}

Both metrics range from 1 to 5. Higher scores indicate better performance.

\section{Results}

\subsection{RQ1: Performance on Code Search}

As the ultimate goal of query reformulation, we first evaluate whether the reformulated queries by \ourmethod lead to better code search performance. 
We experiment under both search engines and compare the improvement of MRR scores before and after query reformulation by various methods. For each query, we calculate its similarity to code instances in the test set. The top 100 instances with the highest similarity are selected as the search results. Each query has one ground-truth code instance in the test set. We calculate the MRR scores by comparing the results and the ground-truth code.
Then, for each method, we select the first three reformulations and report the highest MRR score among them. 
Since the purpose of query reformulation is to hit the potential search intent of the user. We believe that results with the maximum MRR in the top-$k$ reformulations are the most likely to satisfy this goal and are therefore considered meaningful.

\begin{table}[!t]
    \caption{\rmfamily Performance of Various Approaches in Code Search}
    \rmfamily
    \label{tab:performance}
    \begin{threeparttable}
    \setlength\tabcolsep{10pt}
    \begin{tabular}{l@{}ll}
    \toprule
    \textbf{Search Engine\;} & \textbf{~~Approach} & \bf ~~~MRR \\
    \bottomrule
    \multirow{6}{*}{\textbf{CodeBERT}} & ~~No Reformulation & 0.202  \\ 
     & ~~Supervised$^1$~\cite{sequer} & 0.222 (+9.90\%)
     \\ \cline{2-3}
    
     & ~~LuSearch~\cite{lusearch} & 0.144 (-28.70\%)  \\
     & ~~NLP2API~\cite{nlp2api} & 0.148 (-26.87\%)  \\
     & ~~\ourmethod (ours) & \textbf{0.222} (+\textbf{9.90\%}) \\
    \bottomrule
    \multirow{6}{*}{\textbf{Lucene}} & ~~No Reformulation & 0.133   \\ 
     & ~~Supervised~\cite{sequer} & \textbf{0.155} (+\textbf{16.91\%}) \\ \cline{2-3}
    
     & ~~LuSearch~\cite{lusearch} & 0.129 (-2.64\%) \\
     & ~~NLP2API~\cite{nlp2api} & 0.136 (+2.87\%) \\
     & ~~\ourmethod (ours) & 0.149 (+12.23\%) \\
    \bottomrule
    \end{tabular}
    \begin{tablenotes}
    \item $^1$ The result is not reproducible due to the unavailability of the confidential parallel data they use.
    \end{tablenotes}
    \end{threeparttable}
    \end{table}

The experimental results are presented in Table \ref{tab:performance}. \ourmethod enhances the MRR by 9.90\% and 12.23\% on the two search engines, respectively. 
Compared to the two unsupervised baselines, LuSearch and NLP2API, it brings a giant leap of over 50\% in search accuracy. 
More surprisingly, \ourmethod achieves competitive results to the supervised counterpart though it is not given with any annotations. 
This indicates that our self-supervised approach can assist developers to write high-quality queries, which ultimately leads to better code search results. 

We notice that the performance of \ourmethod is slightly worse than the supervised counterpart with the Lucene search engine. This is probably because the supervised approach applies fixed expansion patterns to queries and therefore tends to expand queries with common, fixed keywords. These keywords can be easily hit by search engines based on keyword matching (e.g., Lucene). On the contrary, \ourmethod does not use fixed expansion patterns and thus has more various keywords. Lucene is not able to perform keyword matching on them. Instead, CodeBERT, which models the semantic relationships between keywords, can understand queries expanded by \ourmethod.

Another interesting point is that LuSearch and NLP2API do not contribute to the CodeBERT-based search engine. This is probably because both approaches append words to the tail of the original query, hence perturbing the semantics of the original query when we use deep learning based search engines such as CodeBERT.



\subsection{RQ2: Qualitative Evaluation}

To evaluate the intrinsic quality of the reformulated queries, we perform a human study with programmers. Four participants from author's institution, but different labs, are recruited through invitations. All participants are postgraduates in the area of software engineering or natural language processing, having over-four-year programming experience. 
We took the first 100 queries from the test set in RQ1, 
and reformulated them using various methods, including SEQUER, LuSearch, NLP2API, \ourmethod, and GooglePS. We assigned 100 search tasks to human annotators using these 100 queries and present the reformulated queries by various approaches. The annotators were asked to search code using Google and provide their ratings (on a scale of 1 to 5) towards the reformulation in terms of informativeness and naturalness, without knowing the source of the reformulation tool. 

Table~\ref{tab:HumanEvaluation} summarizes the quality ratings by annotators. Overall, \ourmethod achieves the most improvement in terms of naturalness (19\%) and informativeness (26\%), showing that it reformulated queries are more human-like. Comparatively, GooglePS and SEQUER have much less improvement. LuSearch and NLP2API even decrease the naturalness and informativeness, as they directly append relevant APIs or synonyms to the tail of the original query, and thus break the coherence of the query.

\begin{table}[!t]
\caption{\rmfamily Human Evaluation Result}
\rmfamily
\centering
\scalebox{1.0}{
\label{tab:HumanEvaluation}
\begin{threeparttable}
\begin{tabular}{lll}
\toprule
\textbf{Approach} & \textbf{Naturalness} & \textbf{Informativeness}   \\ \bottomrule
No Reformulation                 & 3.21 & 3.15 \\
Supervised~\cite{sequer}                 & 3.63 (+13.08\%) 
& 3.44 (+9.21\%) \\
\hline
LuSearch~\cite{lusearch}                & 2.63 (-18.07\%) & 3.17 (+0.63\%) \\
NLP2API~\cite{nlp2api}                & 2.80 (-12.77\%) & 3.50 (+11.11\%) \\
GooglePS~\cite{GooglePS}     &   3.27 (+1.87\%) &  3.33 (+5.71\%) \\
\ourmethod (ours)                & \textbf{3.83} (+\textbf{19.31\%}) & \textbf{3.98} (+\textbf{26.35\%}) \\ \bottomrule
\end{tabular}
\end{threeparttable}
}
\vspace{-3mm}
\end{table}

In particular, compared with the strong baseline SEQUER, \ourmethod obtains a greater improvement of 17.14\% in terms of informativeness, while outperforming slightly in terms of naturalness.
The main reason could be that SEQUER applies three reformulation patterns, \ie, deleting unimportant words, rewriting typos, and adding keywords, where only the last pattern increases the informativeness of the query.
Besides, SEQUER often adds keywords in a monotonous pattern, such as appending ``in Java'' at the tail of the queries; meanwhile, our method can generate diverse spans at the proper positions of the original queries. 
Consequently, the reformulated queries by our approach are more informative.


\subsection{RQ3: Performance under Different Configs}

In this experiment, we evaluate the performance of \ourmethod in code search under different configurations with the CodeBERT search engine. We vary the positioning strategy and the number of candidate positions in order to search for the optimal configuration. 

\smallskip \textbf{Positioning Strategies}. 
Selecting expansion positions is critical to the performance. We compare three strategies, including the entropy-based criterion:
\begin{itemize}
\item RAND randomly selects \textit{k} positions in the original query for expansion.
\item PROB selects the top-\textit{k} positions that have  the maximum probability while predicting their missing content.
\item ENTR selects the top-\textit{k} positions that have the  minimum entropy while predicting their missing content.
\end{itemize}

The results are shown in Table \ref{tab:ablation}. The PROB and ENTR strategies bring a large improvement (around 10\%) to the code search performance. This indicates that both criteria correctly quantify the missing information at various positions. Between these two strategies, ENTR performs slightly better than PROB, probably because ENTR considers the entire distribution of the prediction while PROB just considers the maximum one. 
As expected, the RAND strategy causes a degradation of 6.68\% in code search performance because it selects expansion positions without any guidance, which results in incorrect or redundant expansions.

\begin{table}[!t]
    \caption{\rmfamily Performance of \ourmethod under Different Positioning Strategies}
    \rmfamily
    \centering
    \scalebox{1.0}{
    \label{tab:ablation}
    \begin{threeparttable}
    \setlength\tabcolsep{20pt}
    \begin{tabular}{ccc}
    \toprule
    \textbf{Strategy} & \textbf{MRR} & \textbf{Improvement}   \\ \bottomrule
    RAND                 & 0.1886 & -6.68\%  \\
    PROB                 & 0.2220 & +9.85\%  \\
    ENTR                 & \textbf{0.2221} & +\textbf{9.90}\%  \\ \bottomrule
    \end{tabular}
    \end{threeparttable}
    }
\end{table}
\begin{table}[!t]
\caption{\rmfamily Performance of \ourmethod under Different Candidate Position Numbers}
\rmfamily
\centering
\scalebox{1.0}{
\label{tab:ablation2}
\begin{threeparttable}
\setlength\tabcolsep{15pt}
\begin{tabular}{ccc}
\toprule
\textbf{\# Positions} & \textbf{MRR} & \textbf{Improvement}   \\ \bottomrule
1                 & 0.1726 & \textminus14.60\% \\
2                 & 0.2074 & +~\,2.62\% \\
3                 & \textbf{0.2221} & +~\,\textbf{9.90}\% \\ \bottomrule
\end{tabular}
\end{threeparttable}
}
\end{table}

\smallskip \textbf{Number of Candidate Positions}. We also investigate how many expansions lead to the best performance. We vary the number of candidate positions from 1 to 3 and verify their effects on performance.
Table \ref{tab:ablation2} shows the results. We observe that increasing the number of candidate positions has a positive effect on performance. 
The best performance is achieved when 3 candidate positions are expanded. Meanwhile, only one candidate position can have a negative effect on query reformulation. 
The reason can be that our method reformulates the original query with a variety of query intents. 
A larger number of candidate positions can hit more user intents and hence leads to better search accuracy. 

\subsection{Qualitative Analysis}
\begin{table*}[!t]
  \caption{\rmfamily Examples of Query Reformulation by Various Methods}
  \rmfamily
  \centering
  \label{tab:casestudy}
  \begin{tabular}{rl}
  \toprule
  \rowcolor[HTML]{EFEFEF} \bf Original & The total CPU load for the Synology DSM   \\       
    \bf SEQUER & The CPU load the Synology DSM \\ 
    \bf LuSearch & The total CPU load for the Synology DSM \textbf{sum total aggreg} \\ 
    \bf NLP2API & The total CPU load for the Synology DSM \textbf{OperatingSystemMXBean ProcCpu String} \\ 
    \bf GooglePS & The total CPU load for the Synology DSM \textbf{7} \\ 
  \bf \ourmethod & The total \textbf{memory usage and} CPU load for the Synology DSM \\
  
   \hline
  \rowcolor[HTML]{EFEFEF} \bf Original & Fetch the events         \\  
  \bf SEQUER & Fetch the events \textbf{c++} \\ 
  \bf LuSearch & Fetch the events \textbf{bring get convei} \\ 
  \bf NLP2API & Fetch the events \textbf{CalendarEntry ManyToOne OneToMany} \\ 
  \bf GooglePS & \textbf{Service worker} fetch event\\ 
  \bf \ourmethod & Fetch the events \textbf{from the server} \\
  
  \hline
  \rowcolor[HTML]{EFEFEF} \bf Original & Get method that raises MissingSetting if the value was unset.  \\
  \bf SEQUER & Get method that raises MissingSetting if the value was unset \textbf{in c\#} \\ 
  \bf LuSearch & Get method that raises MissingSetting if the value was unset. \textbf{beget get engend} \\ 
  \bf NLP2API & Get method that raises MissingSetting if the value was unset. \textbf{Praveen Kumar Date} \\ 
  \bf GooglePS & \textbf{PHP foreach} unset. \\ 
  \bf \ourmethod  & Get method that raises \textbf{an exception with} MissingSetting if the value was unset. \\
  
  \hline
  \rowcolor[HTML]{EFEFEF}
  \bf Original & Load values from a dictionary structure. Nesting can be used to represent namespaces.  \\
  \bf SEQUER & Load values from a dictionary structure. Nesting can be used to represent namespaces. \\ 
  \bf LuSearch & Load values from a dictionary structure. Nesting can be used to represent namespaces. \textbf{cargo lade freight} \\ 
  \bf NLP2API & Load values from a dictionary structure. Nesting can be used to represent namespaces. \textbf{Amino TKey TValue} \\ 
  \bf GooglePS & Python namespace class \\ 
  \bf \ourmethod  & Load \textbf{a list of} values from a dictionary structure. Nesting can be used to represent namespaces. \\
  
  \bottomrule
  \end{tabular}
  \end{table*}

To further understand the capability of \ourmethod, we qualitatively examine the reformulation samples by various methods. Four examples are provided in Table~\ref{tab:casestudy}.
Example 1 compares the reformulation for the query ``The total CPU load for the Synology DSM'' by various methods. The original query aims to find the code that monitors the CPU load of a DSM. The reformulated query by \ourmethod is more precise to the real scenario since CPU load and memory usage are often important indicators that need to be monitored simultaneously. In contrast, SEQUER only removes the ``total'' at the beginning of the query during reformulation. LuSearch appends the query with the synonyms of the keyword ``total'' such as ``sum'' and ``aggreg''. Meanwhile, NLP2API appends the query with APIs that are relevant to CPU and operating system, which are more useful compared to those of SEQUER and LuSearch. GooglePS appends the word ``7'' after ``DSM'' to indicate the version of DSM, which helps to narrow the range of possible solutions.

In Example 2, the original query ``Fetch the events'' is incomplete and ambiguous because the user does not specify what events to fetch and where to fetch them from. The reformulated query by \ourmethod is more informative than that by SEQUER: \ourmethod specifies the source of events, i.e., from the server, which makes the query more concrete and understandable; meanwhile, SEQUER only restricts the programming language of the target code, without alleviating the ambiguity of the original query. LuSearch expands the synonyms of ``Fetch'' such as ``get'' and ``convei'' to the tail of the query. NLP2API adds APIs relevant to events to the original query. But these synonyms and APIs have limited effect on improving search accuracy. GooglePS specifies the requirement of the query to be a service by adding ``Service worker'' at the beginning of the original query. But such a specification has a limited effect on narrowing the search space.

Example 3 shows the results for the query ``Get method that raises MissingSetting if the value was unset.'' The reformulated query by \ourmethod recognizes that MissingSetting is an exception and prepends it with the exception keyword. This facilitates the search engine to find code with similar functionality. In contrast, SEQUER just specifies the programming language of the target code. Compared to \ourmethod and SEQUER, LuSearch and NLP2API only append irrelevant APIs and synonyms to the original query. Hence, the semantics of the query are broken. GooglePS fails to reformulate such a long query. Instead, it returns a search query from other users that contains the keyword ``unset''. The returned results by GooglePS discard much information from the original query, making  deviate from the user intent.


Finally, the last example shows a worse case.
Although \ourmethod achieves the new state-of-the-art, it might occasionally produce error reformulations.
\ourmethod prepends a modifier ``a list of'' in front of the word ``values'', which conflicts with ``dictionary'' in the given query and thus hampers the code search performance. This is probably because the word ``values'' occurs frequently in the training corpus and often refers to elements in arrays and lists. Therefore, \ourmethod tends to expand it with modifiers such as ``all the'' and ``a list of''. Comparably, SEQUER does nothing to the original query. LuSearch concerns ``Load'' as the keyword and expands it. NLP2API adds APIs relevant to the key and value of the dictionary data structure, which results in better search performance. GooglePS cannot handle such a long query and just gives an irrelevant reformulation.

These examples demonstrate the superiority of \ourmethod in query reformulation for code search, affirming the strong ability of both position prediction and span generation.
In future work, we will conduct empirical research on the error types, and improve our model for the challenging reformulations.

\section{Discussion}
\subsection{Strength of \ourmethod over fully supervised approaches?}

One debatable question is what are the benefits of \ourmethod since it does not beat the SOTA fully-supervised approach in terms of the code search metrics. 


Fully-supervised methods such as SEQUER achieve the state-of-the-art performance by sequence-to-sequence learning on a parallel query set. However, acquiring such parallel queries is infeasible since the query evolution log by search engines such as Google and Stack Overflow is not publicly available. Besides, the sequence-to-sequence approach tends to learn generic reformulation patterns, \eg, specifying the programming language or deleting a few irrelevant words.

Compared to SEQUER, \ourmethod does not rely on the supervision of parallel queries, instead, it is trained on a nonparallel dataset (queries only) that does not need to collect the ground-truth reformulations. This significantly scales up the size of training data, and therefore allows the model to learn diverse reformulation patterns from a large number of code search queries. 
\ourmethod provides an alternative feasible and cheap way of achieving the same performance. 

\subsection{Limitations and Threats}
We have identified the following limitations and threats to our method:


\smallskip\emph{Patterns of query reformulation.} In this work, we mainly explore query expansion, the most typical class of query reformulation. While query expansion is only designed to supplement queries with more information, redundant or misspelled words in the query can also hamper the code search performance, which cannot be handled by our method. Thus, in future work, we will extend our approach to support more reformulating patterns, including query simplification and modification. For example, in addition to only inserting a \verb|[MASK]| token in the CQC task, we can also replace the original words with a \verb|[MASK]| token or simply delete a token and ask the pre-trained model to predict the deletion position. A classification model can also be employed to decide whether to add, delete or modify keywords in the original query.

\smallskip\emph{Code comments as queries.} As obtaining real code queries from search websites is difficult, we use code comments from code search datasets to approximate code queries in building and evaluating our model. Although code comments are widely used for training machine learning models on NL-PL matching~\cite{deepcs,feng2020codebert,wang2021codet5}, they may not represent the performance of queries in real-world code search engines.



\section{Related Work}
\subsection{Query Reformulation for Code Search}
Query reformulation for code search has gained much attention in recent years~\cite{sesimilarwords, co-occurrence, SWordNet, nlp2api, pseudo, lusearch}. There are approximately three categories of technologies, namely, knowledge-based, feedback-based, and deep learning based approaches.

The knowledge-based approaches aim to expand or revise the initial query based on external knowledge such as WordNet~\cite{WordNet} and thesauri. 
For example, Howard \etal~\cite{sesimilarwords} reformulated queries using semantically similar words mined from method signatures and corresponding comments in the source code. Satter and Sakib~\cite{co-occurrence} proposed to expand queries with co-occurring words in past queries mined from code search logs. Yang and Tan~\cite{SWordNet} constructed a software-specific thesaurus named SWordNet by mining code- comment mappings. They expanded queries with similar words in the thesaurus. Lu \etal~\cite{lusearch} proposed LuSearch which extends queries with synonyms generated from WordNet.

Unlike knowledge-based approaches, feedback-based approaches identify the possible intentions of the user from the initial search results and use them to update the original query. 
For example, Rahman and Roy~\cite{nlp2api} proposed to search Stack Overflow posts using pseudo-relevance feedback. Their approach identifies important API classes from code snippets in the posts using TF-IDF, and then uses the top-ranked API classes to expand the original queries. Hill \etal~\cite{pseudo} presented a novel approach to extract natural language phrases from source code identifiers and hierarchically classify phrases and search results, which helps developers quickly identify relevant program elements for investigation or identify alternative words for query reformulation.

Recently, deep learning has advanced query reformulation significantly~\cite{sequer, seq2seqOfBugReport}. Researchers regard query reformulation as a machine translation task and employ neural sequence-to-sequence models. 
For example, Cao \etal~\cite{sequer} trained a sequence-to-sequence model with an attention mechanism on a parallel corpus of original and reformulated queries. The trained model can be used to reformulate a new query from Stack Overflow. 

While deep learning based approaches show more promising results than previous approaches, they rely on the availability of large, high-quality query pairs. For example, Cao \etal's work requires the availability of query pairs within the same session in the search logs of Stack Overflow. But such logs are confidential and unavailable to researchers. This restricts their practicality in real-world code search.

Unlike these works, \ourmethod is a data-driven approach based on self-supervised learning. \ourmethod expands queries by pre-training a Transformer model with corrupt query completion on large unlabeled data. Results demonstrate that \ourmethod achieves competitive results to that of fully-supervised models without requiring data labeling.

\subsection{Code Intelligence with Pre-trained Language Models}

In recent years, there is an emerging trend in applying pre-trained language models to code intelligence \cite{feng2020codebert,wang2021codet5,niu2022spt,hadi2022effectiveness}. 
For example, Feng \etal\cite{feng2020codebert} pre-trained the CodeBERT model based on the Transformer architecture using programming and natural languages. 
CodeBERT can learn the generic representations of both natural and programming languages that can broadly support NL-PL comprehension tasks (\eg, code defect detection, and natural language code search) and generation tasks (\eg, code comment generation, and code translation). Wang \etal~\cite{wang2021codet5} proposed CodeT5, which extends the T5 with an identifier-aware pre-training task. Unlike encoder-only CodeBERT, CodeT5 is built upon a Transformer encoder-decoder model. It achieves state-of-the-art performance on both code comprehension and generation tasks in all directions, including PL-NL, NL-PL, and PL-PL. 

To the best of our knowledge, \ourmethod is the first attempt to apply PLM in query reformulation, which aims to leverage the knowledge learned by PLM to expand queries.

\section{Conclusion}
In this paper, we propose \ourmethod, a novel self-supervised approach for query reformulation. 
\ourmethod formulates query expansion as a masked query completion task and pre-trains T5 to learn general knowledge from large unlabeled query corpora. For a search query, \ourmethod guides T5 through enumerating multiple positions for expansion and selecting positions that have the best information gain for expansion.
We perform both automatic and human evaluations to verify the effectiveness of \ourmethod. 
The results show that \ourmethod generates useful and natural-sounding reformulated queries, outperforming baselines by a remarkable margin.
In the future, we will explore other reformulation patterns such as query simplification and modification besides query expansion. 
We also plan to compare the performance of our approach with large language models such as GPT-4.

\section*{Data Availability}
Our source code and experimental data are publicly available at \href{https://github.com/RedSmallPanda/SSQR}{https://github.com/RedSmallPanda/SSQR}.

\section*{Acknowledgments}
This research is supported by National Natural Science Foundation of China (Grant No. 62232003, 62102244, 62032004) and CCF-Tencent Open Research Fund (RAGR20220129). 

\balance
\bibliographystyle{ACM-Reference-Format}
\bibliography{ref}

\end{document}